\pdfoutput=1
\documentclass[%
twocolumn,
superscriptaddress,
nofootinbib,
 amsmath,amssymb,
 aps, prd
]{revtex4-2}

\usepackage{xcolor} 
\usepackage[normalem]{ulem} 
\usepackage{lipsum} 
\usepackage{soul}
\usepackage{ulem}
\newcommand{\ye}[0]{$Y_e$}

\usepackage{mathtools}
\usepackage{graphicx}
\usepackage{amsmath}
\usepackage{hyperref}

\DeclareSymbolFont{cmletters}{OML}{cmm}{m}{it}
\DeclareMathSymbol{v}{\mathalpha}{cmletters}{"76}

\renewcommand{\textcolor}[2]{#2}

\begin{document}

\title{Black Hole-Neutron Star Binaries near Neutron Star Disruption Limit in the Mass Regime of Event GW230529}

	\author{Tia Martineau}
        \email[Corresponding author: ]{tmartineau1@une.edu}
	\affiliation{Department of Physics and Astronomy, University of New Hampshire, 9 Library Way, Durham, NH 03824, USA}

    \author{Francois Foucart}
	\affiliation{Department of Physics and Astronomy, University of New Hampshire, 9 Library Way, Durham, NH 03824, USA}

    \author{Mark A. Scheel}
    \affiliation{TAPIR, Walter Burke Institute for Theoretical Physics, MC 350-17, California Institute of Technology, Pasadena, California 91125, USA}

    \author{Matthew D. Duez}
    \affiliation{Department of Physics \& Astronomy, Washington State University, Pullman, Washington 99164, USA}

    \author{Lawrence E. Kidder}
    \affiliation{Cornell Center for Astrophysics and Planetary Science, Cornell University, Ithaca, New York, 14853, USA}

    \author{Harald P. Pfeiffer}
    \affiliation{Max Planck Institute for Gravitational Physics (Albert Einstein Institute), D-14467 Potsdam, Germany}

\begin{abstract}

In May 2023, the LIGO–Virgo–KAGRA (LVK) Collaboration reported the likely black hole–neutron star (BHNS) merger GW230529\_181500. The signal was observed with high significance in only one detector, limiting constraints on the black hole spin and motivating our study of disruption in this mass regime. That event is expected to be the merger of a 2.5-4.5 $M_{\odot}$ primary with a secondary compact object of mass between 1.2-2.0 $M_{\odot}$. This makes it the first BHNS merger with a significant potential for the production of electromagnetic (EM) counterparts, and provides further evidence for compact objects existing within the suspected lower mass gap. To produce post-merger EM transients, the component of the black hole spin aligned with the orbital angular momentum must be sufficiently high, allowing the neutron star to be tidally disrupted. The disrupting BHNS binary may then eject a few percent of a solar mass of matter, leading to an observable kilonova driven by radioactive decays in ejecta, and/or a compact-binary GRB (cbGRB) resulting from the formation of an accretion disk and relativistic jet. Determining which mergers lead to disruption of the neutron star is necessary to predict the prevalence of EM signals from BHNS mergers, yet most BHNS simulations so far have been performed far from the minimum spin required for tidal disruption. Here, we use the Spectral Einstein Code (SpEC) to explore the behavior of BHNS mergers in a mass range consistent with GW230529\_181500 close to that critical spin, and compare our results against the mass remnant model currently used by the LVK Collaboration to predict the probability of tidal disruption. Our numerical results reveal the emergence of non-zero accretion disks even below the predicted NS disruption limit, of low mass but capable of powering cbGRBs. Our results also demonstrate that the remnant mass model underpredicts the disk mass for the DD2 equation of state (EOS), while they are within expected modeling errors for SFHo. The disruption limit itself, however, is not found to significantly differ from the predictions of the analytical model, unless remnant masses $M_{\rm rem}\lesssim 0.001M_\odot$ prove interesting observationally. In all of our simulations, any kilonova signal would be dim and most likely dominated by post-merger disk outflows.
  
\end{abstract}

\maketitle

\section{Introduction}\label{sec:introduction}

Until the announcement of gravitational wave event GW230529\_181500  (herein abbreviated GW230529) \citep{Abac_2024}, the detection of a black hole-neutron star (BHNS) binary merger capable of producing electromagnetic (EM) signals had yet to be made by ground-based gravitational wave (GW) detectors. Tidal disruption of the neutron star in BHNS binaries leads to the production of EM counterparts which fall across the entire EM spectrum, including UV/optical/infrared kilonova emission powered by $r$-process nucleosynthesis \citep{Metzger_2019,Gottlieb_2023}, prompt gamma-ray bursts (GRBs) originating from ultra-relativistic jets \citep{1973ApJ...182L..85K, 1991AcA....41..257P}, and longer-lasting GRB afterglows produced by the interactions of those jets with the surrounding medium \citep{NAKAR2007166,annurev:/content/journals/10.1146/annurev-astro-081913-035926}. The observation of these types of EM radiation resulting from BHNS binary mergers has been considered to be one of the most anticipated discoveries in this new era of GW astronomy.

When a neutron star undergoes tidal disruption, bound matter from the disrupted neutron star begins to form a torus around the black hole. Generally, material from the disrupted star will remain outside the black hole as an accretion disk, a tidal tail, and/or unbound ejecta. The disk first circularizes and heats through hydrodynamical shocks and then through magnetically-driven turbulence; it is initially cooled by neutrino emission, until the density becomes low enough for the disk to become radiatively inefficient. The magnetorotational instability (MRI) leads to an increase in the strength of the magnetic field, angular momentum transport and heating in the disk, accretion of matter onto the black hole, and the production of mildly relativistic outflows for multiple seconds after the merger \citep{Fern_ndez_2013,Siegel_2017}.
Intermittent (or sometimes nearly continuous) relativistic jets may also be produced in the $\sim$0.1 -- 1  s window of time after merger \citep{PhysRevD.86.084026, Paschalidis_2015, Ruiz_2018, PhysRevD.106.023008, PhysRevD.107.123001, gottlieb2023largescale}.

Before GW230529, it appeared that a majority of BHNS binaries likely involved high-mass and/or low-spin black holes \citep{10.3389/fspas.2020.00046}. Often, this means the neutron star will plunge whole into the black hole rather than disrupt, suppressing the emission of detectable post-merger EM signals. This has been seen in all of the detections of BHNS binaries prior to GW230529 by the LIGO/Virgo/KAGRA (LVK) Collaboration: the neutron star behaves like a point particle throughout the merger, and the coalescence is indistinguishable from that of (highly asymmetric) binary black holes \citep{Foucart:2013psa} since the entire neutron star is very rapidly accreted onto the black hole. In these cases, pre-merger electromagnetic emission is still possible e.g. due to crust shattering \citep{PhysRevLett.108.011102}, magnetospheric activities \citep{Most_2023}, or charged black holes; such signals are however harder to detect than GRBs or kilonovae.

For systems in which the neutron star undergoes disruption, the disruption is intimately connected to the process of mass ejection \citep{Kyutoku_2021}. There are two main components of neutron-rich matter which can be ejected. One is the matter ejected on dynamical timescales (typically milliseconds) during the merger, which is referred to as merger ejecta or dynamical ejecta. The other is unbound from the merger remnant disk by magnetically-driven, neutrino-driven and/or viscosity-driven winds, referred to as disk wind or secular ejecta.

The presence of merger ejecta (or dynamical ejecta) in BHNS mergers has been predicted by theoretical models and numerical simulations \citep{1976ApJ...210..549L,Kyutoku_2013, PhysRevD.87.084006, PhysRevD.90.024026, Kyutoku_2015,PhysRevD.92.024014}. Matter is expelled from the system as the star is tidally disrupted, with most of the ejecta concentrated close to the orbital plane. The characteristics of the dynamical ejecta are influenced by a range of factors, including the mass ratio of the binary, the compactness of the neutron star, the spin of the black hole, and the relative orientation of that spin and of the binary's orbital plane. Anisotropic mass ejection as dynamical ejecta may induce potentially-observable EM emission \citep{Kyutoku_2013}. For aligned black hole spins, the results of Chen {\it et al} \cite{Chen:2024ogz} in fact indicate that the outcome of the merger is largely independent of parameters other than those previously mentioned, including the total mass of the system (but note that at constant mass ratio and for a given equation of state, increasing the total mass of the system typically results in a more compact -- and thus harder to disrupt -- neutron star).

Ejecta associated with the disk wind is generally slower than the dynamical ejecta, and is comprised of magnetically-driven outflows, neutrino driven outflows, and viscous outflows. Disk outflows produced by each of these various respective processes likely differ in velocity, temperature, and composition \cite{Kyutoku_2021}. Current simulations indicate that anywhere between $5\%$ to $40\%$ of the disk mass is eventually ejected rather than accreted \cite{Fernandez:2018kax, Christie_2019,Fernandez:2020oow, gottlieb2023largescale}.

The tidal disruption of a neutron star is required to occur outside the innermost stable circular orbit (ISCO) of the black hole for inducing the aforementioned astrophysically interesting outcomes like the production of EM counterparts during or after merger. It is these processes in particular which may teach us about the structure and behavior of neutron stars in mixed binaries. If we treat the neutron star as a test mass, and set the black hole spin so that it is aligned with the orbital angular momentum of the binary, the ISCO radius goes as $R_{\rm ISCO}$ = $f(\chi_{\rm BH})GM_{\rm BH}/c^2$. Here, \( f \) is a function ranging from 1 to 9 that decreases as the black hole spin \( \chi_{\mathrm{BH}} \) increases, with \( 0 \le \chi_{\mathrm{BH}} \le 1 \) for prograde rotation \citep{Bardeen1972fi,10.3389/fspas.2020.00046}. For large mass ratios and in Newtonian physics, the disruption radius scales as $R_{\rm dis} \sim k(M_{\rm BH}/M_{\rm NS})^{1/3}R_{\rm NS}$. $k$ is a constant with dependence on the EOS and the black hole spin \citep{1973ApJ...185...43F,Wiggins_2000,10.3389/fspas.2020.00046}. From these relationships, it is evident disruption is more likely for BHNS binaries with low-mass black holes, large prograde black hole spins, and/or large neutron star radii.

Models used to predict the remnant disk masses from disrupting BHNS binary mergers have been built based on known results from numerical simulations. Foucart et al. 2012 (FF12) \cite{Foucart_2012}, for example, was calibrated using 31 different BHNS simulations performed with SpEC \cite{PhysRevD.85.044015}, SACRA \cite{PhysRevD.84.064018}, and the UIUC numerical relativity code \cite{Etienne_2009}. These simulations were mostly performed using simple polytropic equations of state and using fairly old numerical codes. While continuing to work well within its expected range of validity (mass ratio \( q \equiv M_{\mathrm{BH}}/M_{\mathrm{NS}} \ge 3 \)), FF12 leads to inaccurate estimates for whether tidal disruption will occur and how much mass remains outside of the black hole when used at lower mass ratio. 

Foucart et al. 2018 (hereafter FNH18) \cite{Foucart_2018} attempted to improve predictions in that low mass ratio regime. FNH18 was calibrated using 75 NR simulations, including the 31 used for FF12. Of the 44 new simulations, 11 lie outside of the expected range of acceptable predictions by the FF12 model (such as in the low mass ratio regime). FNH18 also includes 12 systems which are described by tabulated and temperature-dependent EOSs. Within that sample, there was no clear deviation between the simulations using tabulated equations of state and those using simpler equations of state models. Even with a more refined remnant mass model, an open problem remains for BHNS mergers: disk mass predictions given by FNH18 have not been systematically tested for spins which lie near the critical BH spin threshold for NS disruption.

\begin{figure}[!h]
    \centering
    \includegraphics[scale=.4]{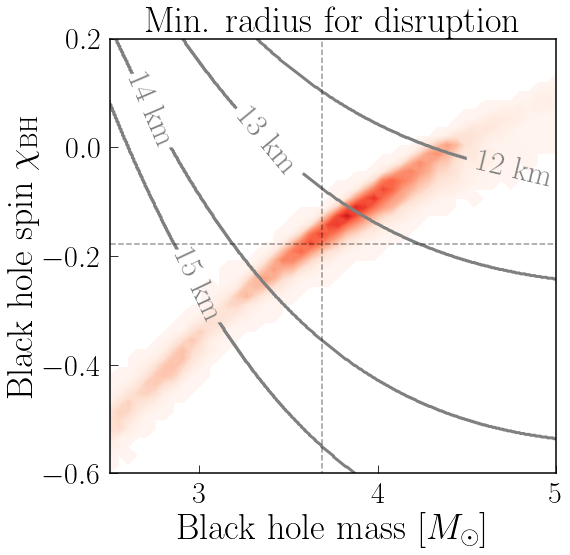}
    \caption{Minimum neutron-star radius (in km) required for tidal disruption of a GW230529-like system, evaluated from the posterior distribution of black-hole masses and spins consistent with LVK results for that event~\cite{Abac_2024}, using the semi-analytical remnant-mass model of Equation~(\ref{mrem})~\cite{Foucart_2018}. Grey contours show the minimum disruption radius, while the red color scale indicates the posterior distribution for GW230529 published by the LVK. When deriving the neutron star mass from the black hole mass, we assume the chirp mass of GW230529. The dashed lines show the mean values of the binary parameters in the posterior. Those values slightly differ from the headline results of~\cite{Abac_2024}, as they quote results using their high-spin prior for the secondary object.}
    \label{fig:gw230529}
\end{figure}

This is particularly important considering one of the most powerful and direct applications of these fitting formulae, i.e. potentially putting direct constraints on the neutron star equation of state from the observation of an EM counterpart to a BHNS merger. This is illustrated in Fig.~\ref{fig:gw230529} for GW230529. In that figure, we show the minimum radius required for tidal disruption of the neutron star across the range of parameters in the publicly available posterior data published by the LVK~\cite{Abac_2024}. Specifically, we consider the results for the `low-spin' prior for the lower mass objects, consistent with the non-spinning neutron stars used in this manuscript. We consider all pairs of black hole mass and black hole spin in the posterior, calculating the neutron star mass from the chirp mass of the system (which is known to $2\%$ accuracy), and use the formula from~\cite{Foucart_2018} to determine the minimum radius for which the mass remaining outside of the black hole post-merger is predicted to be non-zero\footnote{This ignores the value of the tidal deformability in the LVK posterior; we consider this a reasonable first-order approximation as GW230529 is uninformative about that tidal deformability. We also note that using the true neutron star mass from the LVK posterior would not meaningfully change these results, given the accuracy with which the chirp mass is known, but it would prevent easy visualization on a 2D plot}. We can see that the observation of an EM counterpart in this system could, in principle, have put stringent constraints on the equation of state of dense matter.  We note that on Fig.~\ref{fig:gw230529} it appears to be more difficult to disrupt the neutron star for lower mass black holes. This occurs here because of the degeneracy between mass ratio and spin (low-mass black holes have counteraligned spin), and, maybe more importantly, because at constant chirp mass low mass black holes are associated with higher mass, and thus more compact, neutron stars. For the black-hole mass of $\sim 4\,M_\odot$ studied here, Fig.~\ref{fig:gw230529} shows that we need a neutron-star radius slightly below $13\,{\rm km}$ for disruption, corresponding to an EOS that can be marginally softer than DD2. Predictions for the amount of mass unbound in this event for stiff EOSs would likely be affected by the discrepancies between the analytical model and simulation results observed in this work.

In this manuscript, we attempt a more careful study of the behavior of BHNS mergers close to the predicted disruption threshold. We fixed the mass ratio at $q=3$, close to the most likely value for GW230529. We consider both the SFHo equation of state (EOS) \citep{Steiner:2013ApJ...774...17S} and the DD2 EOS \citep{PhysRevC.81.015803}, in order to study neutron stars of different compactness, and fixed the neutron star mass $M_{\rm NS}=1.35M_\odot$. This uniquely fixed the predicted disruption limits in the FNH18 model; we then perform simulations for spins close to that limit, in order to test the reliability of the model and the behavior of BHNS mergers close to the tidal disruption limit. These simulations complement the many existing results with sparser sampling of the spin close to the disruption limit but better coverage of strongly disrupting systems existing in the literature (see \cite{10.3389/fspas.2020.00046,Kyutoku_2021,Duez:2024rnv} for recent reviews, as well as \citep{Tsao:2024tme,Izquierdo:2024rbb,Chen:2024ogz} for recently produced BHNS simulations at similar mass ratios). Concurrently, Matur et al. \citep{Matur2025SpinBHNS} present an independent low–mass-ratio scan targeted to GW230529-like systems ($q = 2.6$, $M_c \simeq 1.91\,M_\odot$, where $M_c$ is the chirp mass), varying $\chi_{\rm BH}$ from 0 to 0.8 and reporting a clear spin dependence of the dynamical/fast ejecta along with a spiral wave–driven ejecta component that grows with spin. We note also that the result of \cite{Chen:2024ogz} indicates that our results likely translate to different neutron star masses, although only for the narrow range of masses for which keeping the compactness of the neutron star constant leaves us with neutron stars of physically acceptable radii (i.e. for high mass neutron stars, one would require simulations with significantly higher compactness in order to get realistic radii).

\section{Methods}

\subsection{Disruption model}

We determine the predicted critical spin for tidal disruption according to FNH18. Whether a neutron star tidally disrupts or plunges whole into the black hole depends on the binary mass ratio \(q=M_{\rm BH}/M_{\rm NS}\) \cite{PhysRevD.87.084006}, the symmetric mass ratio \(\eta=q/(1+q)^2\), the black hole spin magnitude \(\chi_{\rm BH}\) \cite{Etienne_2009}, and the neutron star compactness \(C_{\rm NS}=GM_{\rm NS}/(R_{\rm NS}c^2)\), which contains the equation of state information through the stellar radius \(R_{\rm NS}\).
The empirical model FNH18 for estimating the normalized remnant mass is
\begin{equation}
\hat{M}^{\rm rem}_{\rm model} =
\left[\mathrm{Max}\!\left(
\alpha\frac{1-2C_{\rm NS}}{\eta^{1/3}}
-\beta\,\hat{R}_{\rm ISCO}\frac{C_{\rm NS}}{\eta}
+\gamma,\,0\right)\right]^{\delta},
\label{mrem}
\end{equation}
\textcolor{blue}{
where \(\hat{R}_{\rm ISCO}\) is the dimensionless ISCO radius (Eq.~\ref{eq:risco}).
This model results in the normalized remnant mass \(\hat{M}^{\rm rem}=M^{\rm rem}/M_{\rm NS}^{\rm b}\), where \(M_{\rm NS}^{\rm b}\) is the baryonic mass of the neutron star. Because of gravitational binding, the gravitational mass of a typical neutron star is about 10\% lower than its baryonic mass.}

For a Kerr black hole \cite{Bardeen1972fi},
\begin{align}
\hat{R}_{\rm ISCO} &= \frac{R_{\rm ISCO}}{M_{\rm BH}}
= 3 + Z_{2} \mp \sqrt{(3 - Z_{1})(3 + Z_{1} + 2Z_{2})},
\label{eq:risco} \\
Z_{1} &= 1 + (1 - \chi_{\rm BH}^{2})^{1/3}
\big[(1 + \chi_{\rm BH})^{1/3} + (1 - \chi_{\rm BH})^{1/3}\big], \\
Z_{2} &= \sqrt{3\chi_{\rm BH}^{2} + Z_{1}^{2}}.
\end{align}

A fit of Eq.~\eqref{mrem} to numerical-relativity results gives
$\alpha = 0.406$, $\beta = 0.139$, $\gamma = 0.255$, and $\delta = 1.761$.
A zero value of $\hat{M}^{\rm rem}_{\rm model}$ corresponds to no tidal disruption of the neutron star,
while positive values indicate that some material remains outside the black hole.
The critical spin for tidal disruption can be estimated from Eq.~\eqref{mrem}
by finding $\chi_{\rm BH}$ for which $\hat{M}^{\rm rem}_{\rm model}=0$;
alternatively, the relation can be inverted to obtain the critical neutron star radius for fixed masses and spins.
Figure~\ref{fig:gw230529} shows this threshold for parameters consistent with GW230529.

The following expression is used as an error estimate for $\hat{M}^{\rm rem}$:
\begin{equation}
\sigma_{\rm model}=1.4\left[\left(\frac{\hat{M}_{\rm rem,model}}{10}\right)^2+\left(\frac{1}{100}\right)^2\right]^{1/2},
\label{eq:sigma}
\end{equation}
which accounts for a relative accuracy of approximately 10\% in the remnant-mass measurements of the numerical simulations,
together with an absolute uncertainty of $\sim0.01\,M_{\rm NS}^{\rm b}$ for small masses.
The factor of 1.4 corrects for the root-mean-square error of the fit to the underlying simulation data \cite{Foucart_2018}.
This represents the expected error within the parameter range in which FNH18 was calibrated;
larger deviations are naturally expected outside that range.

\subsection{Numerical methods}

In this study, we model BHNS binaries via numerical simulations using the Spectral Einstein Code (SpEC) \citep{speccode}. SpEC is a fully general relativistic hydrodynamics code that evolves the generalized harmonic~\cite{Lindblom:2005qh} formulation of Einstein’s equations using a pseudospectral method, and the equations of hydrodynamics via shock-capturing, finite-difference techniques~\cite{Duez_2008,PhysRevD.87.084006}. 

Initial conditions for all simulations are obtained using SpEC's internal initial data solver, Spells \cite{Pfeiffer:2005zm,Pfeiffer_2003,Foucart_2008}. Spells solves for the constraints in Einstein’s equations. Additionally, it solves for an irrotational velocity profile inside the neutron star while requiring hydrostatic equilibrium \cite{PhysRevD.63.064029,Taniguchi_2006}. When initially solving Einstein's equations, SpEC sets the radial velocity of the stars to zero, but in reality the stars follow a spiral trajectory, and ignoring this results in eccentricity. The trajectory's eccentricity is reduced after a short 3-orbit evolution of the binary according to an iterative procedure described in Buonanno {\it et al}~\cite{Buonanno_2011}.

The evolution of the binary is performed using a two-grid method. There is a pseudo-spectral grid for evolving Einstein's equations, and a separate Cartesian grid utilizing finite difference methods for general relativistic fluid evolution. SpEC uses the WENO5 reconstructor \cite{Jiang1996202,Borges} with an Harten-Lax-van Leer (HLL) approximate Riemann solver~\cite{HLL} for high-order shock-capturing methods. A sphere of constant grid-frame radius is removed from the grid to avoid evolving the interior of the black hole. These simulations did not include magnetic fields or momentum transport to model subgrid magnetic turbulence, as these effects operate on post-merger timescales longer than those covered by our simulations.

Each simulation presented in this manuscript was run with Monte Carlo neutrino radiation transport, following the methods of \cite{Foucart_2021b}. This marks the first use of this method in black hole-neutron star (BHNS) merger simulations. Unlike approximate schemes like leakage models or moment-based approaches, the Monte Carlo method directly simulates neutrino transport by randomly sampling neutrinos and tracking their trajectories in both position and momentum space. This enables accurate modeling of neutrino interactions, including absorption, emission, and scattering, in a highly dynamic merger environment.
The Monte Carlo approach is particularly suited for BHNS mergers, because black hole-disk systems are significantly cheaper to evolve with Monte-Carlo methods than systems in which a dense, hot neutron star remains present. 

Neutrino interactions play a critical role in determining the thermal evolution and composition of the merger ejecta, but generally do not impact the dynamics of BHNS mergers. Here, Monte-Carlo transport is thus mainly useful in order to capture the composition of the post-merger remnant, an important step if we wish to understand outflow properties and potential $r$-process nucleosynthesis. By capturing these effects with higher fidelity than traditional schemes, this method offers more reliable insights into how neutrino radiation influences the observable properties of BHNS mergers, including potential electromagnetic counterparts like kilonovae. We note that even over the relatively short timescales simulated here ($\sim 10\,{\rm ms}$ post-merger), neutrino-matter interactions play a role in black hole-neutron star mergers. The cooling timescale of the remnant due to neutrino-matter interactions is significantly longer than our simulation time, but the timescale over which the composition of the hotter outflows and remnant disk changes is much shorter (as was already known from simulations using more approximate transport schemes, e.g.~\cite{Foucart:2015vpa}), because both the ejecta and the disk are far from their equilibrium composition.

For configurations run at 3 resolutions, initial grid spacings on the finite difference grid are $\Delta$x$^0_{FD}$ = (245.12m, 197.87m, 147.12m) for low, mid-level and high resolution runs, respectively. Simulations performed at a single resolution use the middle resolution. During the evolution, as the grid contracts, the grid spacing decreases by up to $20\%$ before we interpolate the evolved variables on a new grid using the original grid spacing. On the pseudospectral grid, we use adaptive p-refinement, with a target relative truncation error of $\epsilon=3\times 10^{-4} (0.8)^{5L}$ and $L=0,1,2$ for the low, medium, and high resolution.

\subsection{Initial Conditions}

We consider initial configurations for the binary summarized in Table \ref{table:1}. All initial conditions are constraint satisfying and were generated using Spells \cite{Pfeiffer:2005zm,Pfeiffer_2003,Foucart_2008}. The first set of simulations models the neutron star using the SFHo EOS. Each configuration in this group of simulations is a system with mass ratio $q =$ 3, an aligned BH spin, and a non-spinning NS.

The initial black-hole spins $\chi_{\rm BH}$ listed in Table~\ref{table:1} are rounded to three significant figures for consistency. 
Trailing zeros are retained where appropriate to make the stated precision explicit (e.g., $0.0870$), and values close to zero may therefore display additional decimal places while still reflecting three significant figures (e.g., $-0.030$). 
The underlying SpEC initial-data values are known to higher precision.
The critical spin for tidal disruption in the case of the SFHo EOS with a NS of mass 1.35 M$_\odot$ and a radius $R_{\rm NS}$ of 11.9 km is $\chi_{\rm crit} = 0.037$. We then chose to simulate binaries whose initial BH spin magnitudes lie just below and above $\chi_{\rm crit}$.

The same exact calculation was carried out for the DD2 EOS to obtain a value for $\chi_{\rm crit}=-0.23$. The change in EOS is represented in our simulations by a change in compactness of the NS: for DD2,  $R_{\rm NS}=13.2\,{\rm km}$. This lower compactness leads to a lower value of the critical spin. In this case, we ran three simulations at a spin slightly above the critical spin.

We note that this places us close to the estimated parameters of GW230529 for the mass ratio and aligned component of the black hole spin. The chirp mass in our simulation is $M_c=1.98M_\odot$, at the upper bound of the $90\%$ confidence interval of the LVK $M_c=1.94^{+0.04}_{-0.04}M_\odot$~\cite{Abac_2024}. The gravitational wave data for GW230529 does not provide any conclusive evidence for or against the presence of a non-aligned component of the spin. In this manuscript, we do not consider precessing systems.
As shown in Fig~\ref{fig:gw230529}, the two equations of state used here bracket the disruption limit predicted by FNH18 for $M_{\rm BH}\approx 4M_\odot$, with DD2 being closer to that predicted limit. The gravitational wave data is however uninformative about the equation of state of neutron stars.

\begin{table*}[t]
\begin{center}
\begin{tabular}{|c|c|c|c|c|c|c|c|c|c|c|c|c|}
\hline
\multicolumn{1}{|c|}{\textbf{Res.}} & \multicolumn{1}{c|}{\textbf{EOS}} & \multicolumn{1}{c|}{\textbf{$\chi_{\rm i,BH}$}} & \multicolumn{1}{c|}{\textbf{$e$}} & \multicolumn{1}{c|}{\textbf{M$_{\rm rem, FNH18}$}} & \multicolumn{1}{c|}{\textbf{M$_{\rm rem, NR}$}} & \multicolumn{1}{c|}{\textbf{M$_{\rm unbound}$}} & \multicolumn{1}{c|}{\textbf{$\chi_{\rm final,BH}$}} & \multicolumn{1}{c|}{\textbf{M$_{\rm final,BH}$}} & \multicolumn{1}{c|}{\textbf{$\langle Y_{\rm e,disk}\rangle$}} & \multicolumn{1}{c|}{\textbf{$\langle Y_{\rm e,outflows}\rangle$}}\\ \hline
\hline
L1 & SFHo & -0.0130 & 0.0039 & 0.00 & 8.03$\times 10^{-04}$ & 2.46$\times10^{-04}$ & 0.55 & 5.26 & 0.15 & 0.29 \\ 
L0* & SFHo & 0.0870 & 0.0016 & 1.25$\times 10^{-03}$ & 6.60$\times 10^{-03}$ & 1.57$\times 10^{-03}$& 0.60 & 5.25 & 0.15 & 0.23\\ 
L1* & SFHo & 0.0870 & 0.0015 & 1.25$\times 10^{-03}$ & 7.48$\times 10^{-03}$ & 1.68$\times 10^{-03}$ & 0.59 & 5.26 & 0.14 & 0.19\\ 
L2* & SFHo & 0.0870 & 0.0014 & 1.25$\times 10^{-03}$ & 7.65$\times 10^{-03}$ & 2.07$\times 10^{-03}$& 0.59 & 5.26 & 0.13 & 0.18 \\ 
L1 & SFHo & 0.137 & 0.0047 & 4.61$\times 10^{-03}$ & 1.25$\times 10^{-02}$ & 1.97$\times 10^{-03}$& 0.62 & 5.26 & 0.12 & 0.17\\  \hline 
L1 & DD2 & -0.180 & 0.0034 & 8.84$\times 10^{-04}$ & 1.45$\times 10^{-02}$ & 2.77$\times 10^{-03}$ & 0.48 & 5.28 & 0.10 & 0.24\\ 
L1 & DD2 & -0.130 & 0.0086 & 3.36$\times 10^{-03}$ & 3.38$\times 10^{-02}$ & 4.37$\times 10^{-03}$ & 0.51 & 5.27 & 0.09 & 0.18 \\ 
L1 & DD2 & -0.030 & 0.0013 & 1.22$\times 10^{-02}$ & 4.88$\times 10^{-02}$ & 4.12$\times 10^{-03}$& 0.54 & 5.25 & 0.08 & 0.09 \\ \hline 

\end{tabular}
\caption{\label{spins}
Parameters from left to right: simulation resolution (L0 being low resolution, L2 being high resolution), chosen EOS, simulated black hole spin $\chi_{\rm BH}$, eccentricity, FNH18’s mass–remnant prediction, the total simulated mass remaining on the grid at late times (both bound and unbound), the total amount of mass on the grid that is unbound, the final simulated black hole spin $\chi_{\rm final,BH}$, the final Christodoulou mass of the black hole, the average electron fraction in the disk $\langle Y_{\rm e,disk}\rangle$, and the average electron fraction in the outflows $\langle Y_{\rm e,outflows}\rangle$. All masses are given in solar masses ($M_\odot$). Simulations marked with an asterisk (*) correspond to the same configuration run at three separate resolutions. All quantities are reported at approximately $10\,\mathrm{ms}$ after merger, corresponding to the nearest available checkpoint in each simulation. Average electron fractions in the disk and outflows are computed as mass-weighted spatial averages over all fluid elements identified as bound and unbound, respectively, at the measurement time.}
\label{table:1}
\label{tab:summary}
\end{center}
\end{table*}

To assess numerical uncertainties, we repeated the SFHo $\chi_{\rm BH}=0.0870$ configuration at three resolutions (L0, L1, L2). The remnant disk mass differs by about $2\%$ between L1 and L2 and by about $15\%$ between L0 and L2. We conservatively adopt $20\%$ error bars for remnant masses throughout, noting however that using a more optimistic value of $\sim (5-10)\%$ and assuming that it applies to all of our configurations would have no impact on our conclusions. For the unbound mass, we observe $\sim 25\%$ variations between resolutions, but without clear convergence between the three resolutions. Previous SpEC simulations with more massive outflows more similar to our DD2 results showed convergence of the unbound mass to $\sim 20\%$ and of other average outflow properties to $\sim 10\%$~\cite{Foucart:2016vxd}; the very close agreement between L0 and L1 in unbound mass in our simulations is likely accidental and indicative that L0 is not yet in the convergence regime for the small amount of mass unbound in that simulation. We note that this is unfortunately common in merger simulations producing relatively low-mass outflows (see e.g. simulations at similar resolutions in~\cite{2013PhRvD..87b4001H}). We do not claim specific error bars for the mass of these outflows and only consider a qualitative description of their properties here. As for the higher mass outflows of~\cite{Foucart:2016vxd}, the global properties of the outflows (e.g. composition) appear more reliable that their mass. This is not overly suprising as the unbound matter is a small fraction of the ejecta, very sensitive to small variations in the amount of energy given to the outflows, while the other properties of the ejecta are fairly similar for unbound and marginally bound matter.

Residual orbital eccentricities range from $e \approx 0.0016$ to $0.0086$ after reduction. At these levels, the resulting phase errors over the final orbits are $\lesssim 0.3$ rad, and we do not expect eccentricity to have a measurable effect on the bound or unbound mass within our numerical uncertainties~\cite{PhysRevD.84.064018}.

\section{Results}
At the outset of our analysis, we define ``remnant disk mass'' as the bound baryonic mass outside the black hole following the tidal disruption of the neutron star. The ``unbound mass'' refers to material identified as unbound in post-processing using the criterion of Ref.~\cite{Foucart_2021a}. The ``total mass left on the grid'' (or ``total remnant mass'') includes both bound and unbound material remaining outside the black hole within the computational domain. Reported values are measured at approximately 10~ms post-merger, consistent with the assumption of the fitting formula FNH18, which is meant to model the sum of the bound and unbound remnant mass. All remnant and ejecta masses quoted throughout this section are taken directly from Table~\ref{tab:summary} and reported with numerical precision consistent with the tabulated values.

For a BHNS system whose NS is described by the SFHo EOS, and has an initial BH spin of $\chi_{\rm BH} = -0.0130$ (below the critical spin for tidal disruption, $\chi_{\rm crit} = 0.037$), the model predicts that no disruption will occur and we will have no remnant mass. Our simulation yields a small but non-zero \textcolor{blue}{total remnant mass} of 8.03$\times 10^{-04}$ M$_\odot$. For an initial BH spin magnitude of $\chi_{\rm BH} = 0.0870$, just above the calculated critical spin $\chi_{\rm crit}$, FNH18 predicts a \textcolor{blue}{total remnant mass} of $M^{\rm rem}_{\rm model} = $1.25$\times 10^{-03}$ M$_\odot$, and numerical results demonstrate a \textcolor{blue}{total remnant mass} of 7.48$\times 10^{-03}$ M$_\odot$. Another higher spin case was also considered, where the initial spin of the BH is $\chi_{\rm BH} = 0.137$. For this spin, still for a system of mass ratio $q = 3$, FNH18 predicts that NS disruption will lead to a \textcolor{blue}{total remnant mass} of $M^{\rm rem}_{\rm model} = $4.61$\times 10^{-03}$ M$_\odot$. Once again, we find that the resulting \textcolor{blue}{total remnant mass} from our simulation is higher, 1.25$\times 10^{-02}$ M$_\odot$. In all of these cases, however, the difference between simulation and fitting formula is within the assumed numerical error of the fit, $\pm$0.014 M$_\odot$.

\begin{figure}[!h]
    \centering
    \includegraphics[scale=.4]{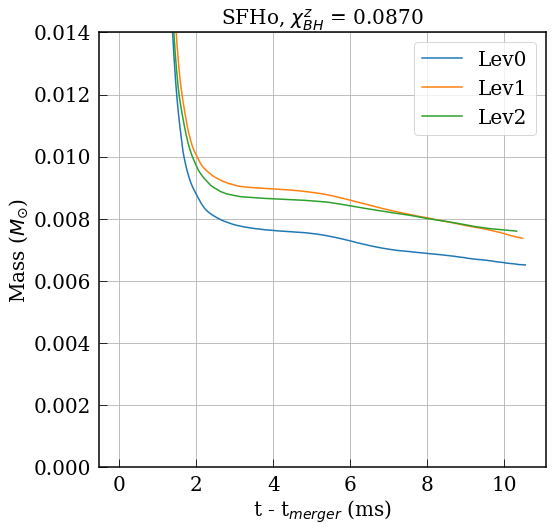}
    \caption{Total remnant mass (bound and unbound) left on the grid for simulations described by the SFHo EOS with $\chi_{\rm BH} = 0.0870$ run at 3 separate resolutions. The maximum relative error between these results is $\sim$20\%.}
    \label{fig:massavgsrescompare}
\end{figure}

To estimate numerical errors in our simulations, we performed a convergence study for the SFHo case with $\chi_{\rm BH} = 0.0870$, running the simulation at three resolutions. We find relative errors of at most $\sim 20\%$ in the resulting \textcolor{blue}{total remnant mass}, which is consistent with prior SpEC studies involving higher-mass remnant disks~\cite{Foucart:2016vxd,Foucart:2019bxj}. Figure~\ref{fig:massavgsrescompare} shows the evolution of the total remnant mass across resolutions, confirming that $20\%$ is a conservative estimate for our standard (Lev1) resolution. Given that SFHo produces lower disk masses—and thus represents a more numerically challenging regime—we adopt this $20\%$ uncertainty as a conservative estimate for all cases, including the higher-mass DD2 configurations, which are expected to be resolved at least as well.

\begin{figure}[!h]
    \centering
    \includegraphics[scale=.4]{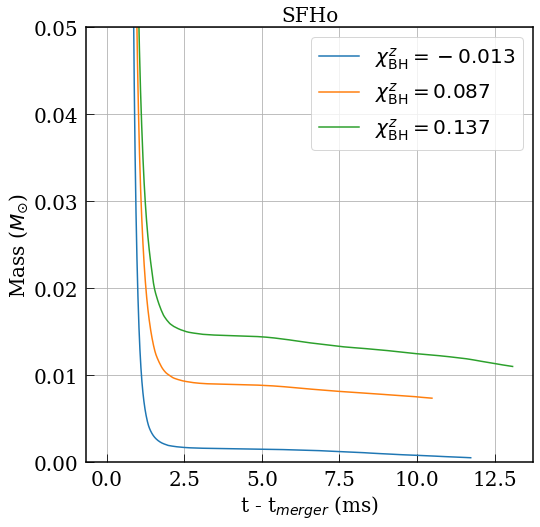}
    \caption{Total mass remnant on the grid for simulations described by the SFHo EOS at least 10 ms after merger.}
    \label{fig:massavgsfho}
\end{figure}

\begin{figure}[!h]
    \centering
    \includegraphics[scale=.4]{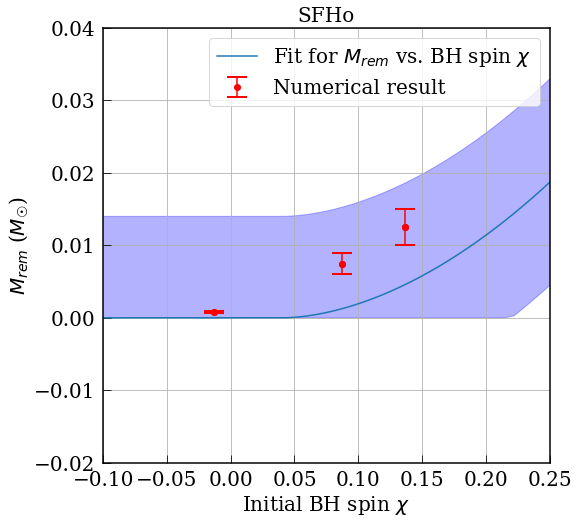}
    \caption{Comparison of numerical results from simulations using SFHo EOS to the predictions of total mass remnant model FNH18.  \textcolor{blue}{The blue shaded region represents the model uncertainty} derived from \ref{eq:sigma}. Error bars on the numerical results represent 20\% of the reported disk mass, based on convergence testing of the SFHo simulations. As the SFHo EOS produces the smallest total remnant masses and is thus the most challenging to resolve, we adopt this 20\% as a conservative, representative estimate of numerical uncertainty for all cases.}
    \label{fig:comparisonsfho}
\end{figure}

The time dependence of the total mass remaining on the grid is shown on Fig.~\ref{fig:massavgsfho}. We see that the total remnant mass mostly stabilizes about $2.5\,{\rm ms}$ post-merger, with later evolution being driven by slow accretion onto the black hole.
As expected, increasing the black hole spin leads to higher total remnant masses. Fig.~\ref{fig:comparisonsfho} compares our simulation results with FNH18. We see that the simulations and model agree within the expected modeling error bars, though the simulations are consistently biased towards higher remnant masses.

\begin{figure}[!h]
    \centering
    \includegraphics[scale=.4]{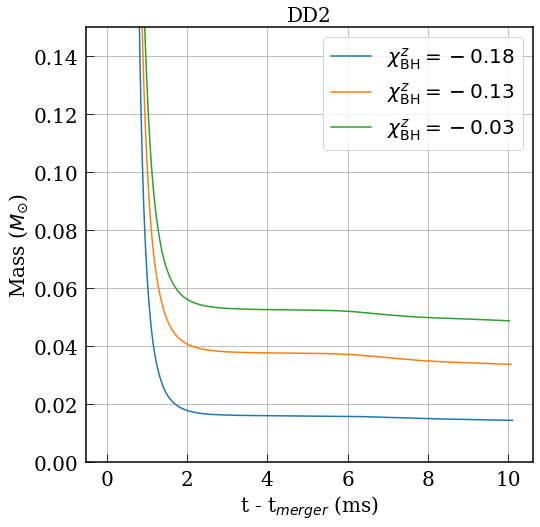}
    \caption{Total mass left on the grid for simulations described by the DD2 EOS at least 10 ms after merger.}
    \label{fig:massavgsdd2}
\end{figure}

\begin{figure}[!h]
    \centering
    \includegraphics[scale=.4]{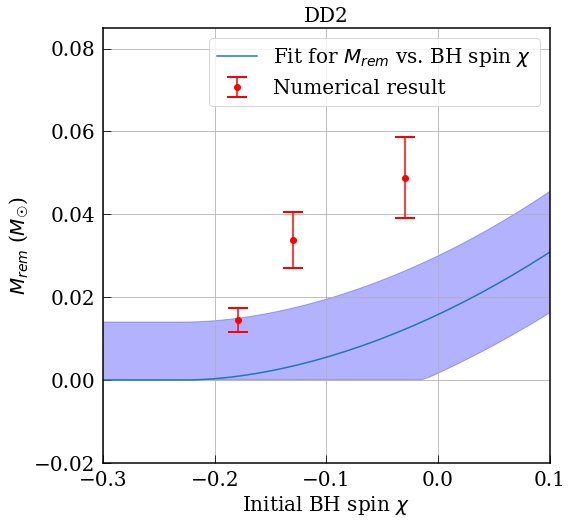}
    \caption{Same as Fig.~\ref{fig:comparisonsfho}, but for the DD2 equation of state}
    \label{fig:comparisonsdd2}
\end{figure}

In contrast to the results we are presenting for the SFHo EOS, the divergence between simulations which use the DD2 EOS and the FNH18 model highlights the sensitivity of the model to the choice of equation of state.
For a BHNS system whose NS is described by the DD2 EOS and a mass ratio of $q =$ 3, we find our critical BH spin threshold for NS disruption to be at $\chi_{\rm crit} = -0.23$. For the DD2 EOS, we consider spins just above this threshold. For a BH of initial spin $\chi_{\rm BH} = -0.180$, the FNH18 model predicts that the binary will produce a total remnant mass of 8.84$\times 10^{-04}$ M$_\odot$. Numerical results demonstrate that the total remnant mass will be significantly higher than this, just barely within the expected range covered by the model's error estimate. The simulation of the merger results in a total remnant mass of 1.45$\times 10^{-02}$ M$_\odot$. Similarly, for an initial BH spin of $\chi_{\rm BH} = -0.130$, FNH18 predicts that the neutron star will disrupt and the total merger remnant will be of mass $M^{rem}_{model} = $3.36$\times 10^{-03}$ M$_\odot$. The numerical results show that this system will produce a \textcolor{blue}{total remnant mass of} 3.38$\times 10^{-02}$ M$_\odot$, lying outside the range of the model and error estimate. We also simulated a BHNS system with a BH spin of $\chi_{\rm BH} = -0.030$. For $\chi_{\rm BH} = -0.030$, FNH18 predicts a \textcolor{blue}{total remnant mass} of $M^{\rm rem}_{\rm model} = $1.22$\times 10^{-02}$ M$_\odot$. The simulated merger results in an overall remnant mass of 4.88$\times 10^{-02}$ M$_\odot$. (See Fig.~\ref{fig:massavgsdd2} for the time evolution of the remnant mass and Fig. \ref{fig:comparisonsdd2} for a visual comparison of numerical results with FNH18.) These results support the correlation between higher initial BH spins and higher total remnant masses, however, two of the three simulations with NSs described by the DD2 EOS have total remnant masses which lie outside of the range of expectation as predicted by FNH18, even after accounting for numerical errors in these simulations and claimed uncertainties in the FNH18 fit. It should be noted particularly in the context of our results for the DD2 EOS that the FNH18 model assumes a mass measurement 10 ms post-merger.

We note however that for both SFHo and DD2, simple extrapolation of the simulations above the critical spin to the point where we would expect $M_{\rm rem}\approx 0$ leads to critical spins in the numerical simulations fairly close to the critical spin predicted by FNH18. \textcolor{blue}{Linear interpolation from the two simulations right above the disruption threshold would lead to differences of $\Delta \chi_{\rm BH} \approx 0.02$ on the critical spin.} 
The main differences between FNH18 and our simulations seem to be that (1) the slope $dM_{\rm rem}/d\chi_{\rm BH}$ close to the critical spin is significantly higher in simulations; and (2) regions of ``zero'' remnant mass in FNH18 may still produce disk masses $\sim 0.001M_\odot$, as seen in the lowest spin SFHo simulation. In e.g. the gamma-ray burst model of Gottlieb {\it et al}~\cite{Gottlieb:2023sja}, this may be (marginally) sufficient to produce a short gamma-ray burst.

In addition to the total amount of matter in the remnant within the domain of our simulations, we consider how much matter in the system is unbound. Although the simulated outflows are modest, their presence may still have implications for the generation of EM counterparts, particularly in the context of prospective kilonovae. We consider as unbound any matter that, at the end of the simulation, satisfies the criteria derived in~\cite{Foucart_2021a}, which takes into consideration the estimated impact after the end of the simulation of both $r$-process heating and neutrino cooling. The amount of unbound matter that leaves the computational domain before the end of the simulation is insignificant ($\mathcal{O}[{10^{-5}}{M_\odot}]$, for our most massive disks). Relative errors on the amount of mass flagged as unbound are typically larger than errors on the disk mass; we observe $\sim 20\%$ error for our 3-resolutions configuration here, below what was observed for more challenging configurations in previous SpEC simulations\citep{Foucart:2016vxd,Foucart:2019bxj}.

Across all SFHo simulations, we find that a maximum of 30\% of all total matter remaining can be categorized as unbound at 10 ms post-merger. For the DD2 simulations, a maximum of 20\% of all of the total remaining matter is unbound, in agreement with the predictions provided in \cite{Chandra:2024ila}. If we assume that, as shown in~\cite{Fernandez:2020oow}, about $20\%$ of the mass remaining in an accretion disk after the merger is ejected through viscous outflows (and an unknown fraction in magnetically-driven winds), we could thus expect a comparable amount of mass to still be unbound over the later evolution of the post-merger remnant. We note that the matter marked as unbound in our simulations $10\,{\rm ms}$ post-merger already includes both cold, neutron rich tidal ejecta, as well as hotter, less neutron rich material ejected from the black hole-disk remnant. Overall, the neutron-rich outflows are thus subdominant compared to the $Y_e\gtrsim 0.2$ outflows.


\begin{figure}[!h]
    \centering
    \begin{minipage}[b]{0.45\textwidth}
        \centering
        \includegraphics[scale=.4]{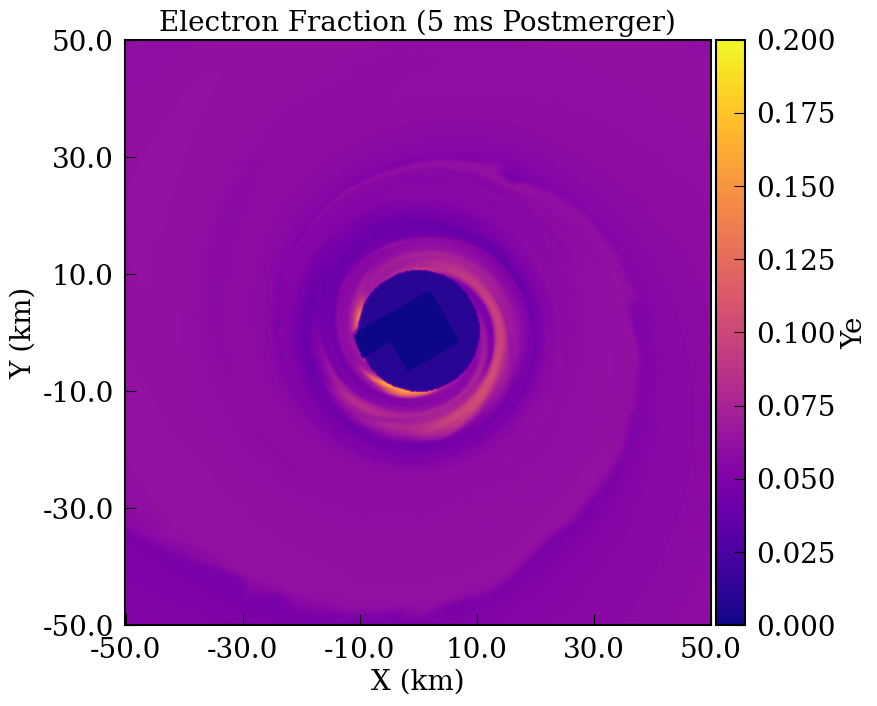}
        \textbf{(a)} Horizontal slice along the equatorial plane of the remnant 5ms postmerger.
        \label{fig:mesh0}
    \end{minipage}
    \hfill
    \begin{minipage}[b]{0.45\textwidth}
        \centering
        \includegraphics[scale=0.4]{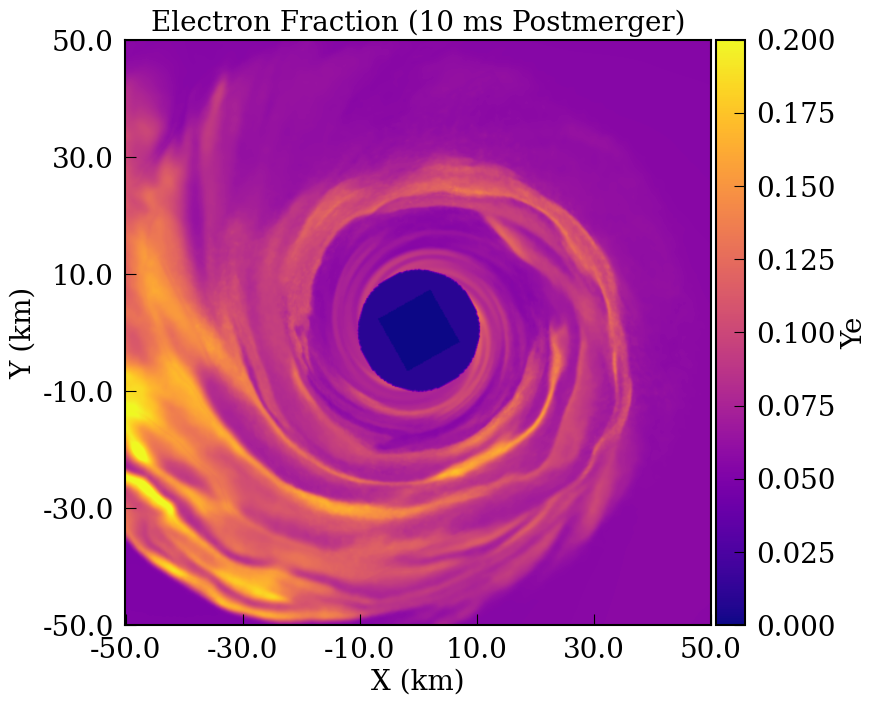}
        \textbf{(b)} Horizontal slice along the equatorial plane of the remnant 10ms postmerger.
        \label{fig:mesh2}
    \end{minipage}
    \caption{Electron fraction ($Y_{e}$) distribution in the equatorial (midplane) slice of the BHNS merger remnant at $5$ and $10\,\mathrm{ms}$ after merger. The color scale shows the instantaneous mass-weighted $Y_{e}$ in the orbital plane. Results correspond to the simulation using the DD2 equation of state with an initial black-hole spin of $\chi_{\rm BH} = -0.0306$. Snapshots are taken from the same simulation outputs used for Table~\ref{spins}, and the slice location is defined by the coordinate midplane of the computational domain.}
    \label{fig:ye_heatmaps}
\end{figure}

Fig.~\ref{fig:ye_heatmaps} presents heatmaps of the electron fraction ($Y_e$) in the equatorial plane of the BHNS merger remnant at 5~ms and 10~ms postmerger. At 5~ms postmerger, the accretion disk is uniformly neutron-rich, with $Y_e$ values predominantly below 0.1 throughout the disk. There is no significant increase in $Y_e$ toward the outer regions, suggesting that any weak interactions have had minimal impact on the electron fraction distribution at this early stage.

By 10~ms postmerger, the overall average $Y_e$ within the disk is already notably higher, though the average electron fraction of the disk remains below 0.2 for all cases ($\langle Y_e \rangle \in [0.08,0.15]$). At 10~ms, the outflows exhibit a broader range of average $Y_e$, $\langle Y_e\rangle \in [0.09,0.29]$. Clearly weak interactions have modified the composition fo the hotter matter ejected from the disk on these timescales. While these changes are not yet substantial enough to significantly alter the neutron-rich character of most of the ejecta, they suggest that weak interactions are gradually influencing the system.

\begin{figure}[!h]
    \centering
    \includegraphics[scale=.45]{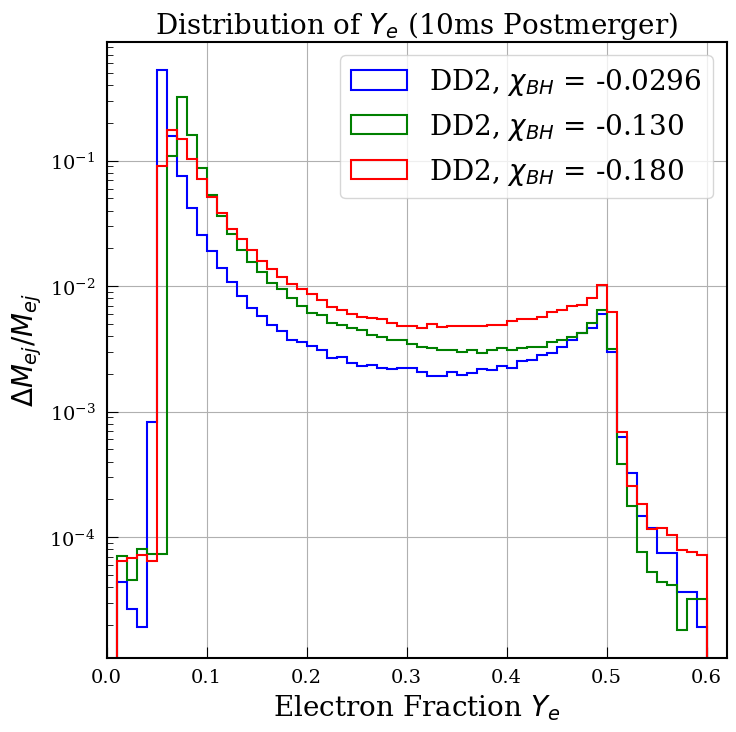}
    \caption{Histograms of the electron fraction \ye of the unbound matter 10 ms postmerger for simulations using the DD2 EoS. The fraction of unbound matter with \ye $<$ 0.1 increases with black hole spin, comprising 75.68\% for the low-spin DD2 case ($\chi_{\rm BH} = -0.180$), 85.26\% for the mid-spin case ($\chi_{\rm BH} = -0.130$), and 87.37\% for the high-spin case ($\chi_{\rm BH} = -0.030$).}
    \label{fig:Ye_dist_dd2}
\end{figure}

The electron fraction distribution of the ejecta, shown in Figure~\ref{fig:Ye_dist_dd2}, indicates that the unbound matter consists primarily of neutron-rich ejecta ($Y_e < 0.1$, from the tidal disruption of the neutron star), along with a smaller fraction of less neutron-rich material ($Y_e \gtrsim 0.2$, mainly ejected at later times). Across all configurations, the DD2 cases exhibit a more neutron-rich composition than the SFHo simulations. However, we should note that the total ejected mass remains low compared to the mass of the bound matter. This suggests that this early ejecta is unlikely to be the dominant contributor to the overall outflows. The higher $Y_e$ outflows already visible at the end of our simulations and the disk wind outflows expected at later times are, when combined, expected to be slightly more massive than the early-time neutron rich outflows observed in our simulations. The difference in composition between equations of state and black hole spins can be largely attributed to differences in the mass of the true ``dynamical ejecta'' in these systems: the DD2 equation of state leads to more ejection of cold matter in a tidal tail, and higher spins have a similar effect.

\section{Discussion}

In this manuscript, we completed a set of six simulations of BHNS mergers with mass ratio $q=3$ and black hole spin (anti)-aligned with the orbital angular momentum. Our simulations, for the SFHo and DD2 equations of state, are meant to study the behavior of these mergers in the poorly studied region of parameter space close to the threshold for tidal disruption. They also happen to match a number of the properties of the recent GW230529 event: similar mass ratio and aligned spin component, although GW230529 allows for the possibility of precessing spins, not considered in this work.

First, in terms of the mass remaining outside of the black hole after merger, a comparison between our simulation results and the commonly used FNH18 total mass remnant model reveals a consistent bias towards predicting lower masses by the model. This bias is particularly evident when considering simulations utilizing the DD2 equation of state, where the numerical results exceed the predictions of FNH18 by more than the expected modeling error. Simulations employing the SFHo equation of state fall within the expected errors of FNH18. 
The critical spin for tidal disruption is however consistent between model and simulations; it is the slope $dM_{\rm rem}/d\chi_{\rm BH}$ that appears inaccurate close to tidal disruption.

We note that the deviations observed for DD2 and, to a lesser extent, SFHo are not unexpected. The FNH18 fitting formula was calibrated primarily on simulations far away from the disruption threshold, and with a fitting method that de-emphasizes capturing the behavior of binaries near the disruption threshold (i.e. assuming errors of at least $0.01M_\odot$ for the mass remaining outside the black hole). Consequently, its accuracy near the disruption threshold for realistic EOSs has not been systematically verified. Our results highlight the limits of the current calibration if BHNS mergers close to the disruption limit can be EM-bright, and thus need to be properly modeled. We also note that the fact that all SFHo simulations agree with the models is in large part due to the fact that the model does not try to capture small remnant masses. All of our SFHo results have remnant mass below the minimum error associated with the model.

The simulation with a black hole spin right around the predicted disruption limit demonstrates the emergence of a very low mass disk with little unbound matter. Interestingly, Gottlieb et al. \citep{Gottlieb:2023sja} have shown that disks of comparable mass can serve as potential progenitors of standard short GRBs -- possibly more easily than the high-mass disks produced by higher spin systems. The high-resolution simulation using an SFHo EOS with a remnant mass of 7.67$\times 10^{-03}$ M$_\odot$ was in fact used as the initial conditions for a long-term disk evolution in~\cite{Gottlieb:2023sja}, with the simulation results corroborating that very low mass disk remnants are viable progenitor candidates for cbGRB production, making low-mass disk outcomes of special interest. We note that if created in such a scenario, a short GRB would be quite different from e.g. the EM counterparts to GW170817: the limited amount of mass ejection would make the detection of an associated kilonova very unlikely. 

Additionally, our results provide information about mass ejection from BHNS binaries near the neutron star disruption limit. Most of the mergers considered here have very little mass ejected during the tidal disruption itself ($\lesssim 0.004M_\odot$). In \cite{Fernandez:2020oow}, the expected mass ejection from viscous disk outflow for mass ratio $q\sim 3$ was shown to be $\sim 20\%$ of the disk mass. More matter may be unbound due to magnetically driven winds for favorable magnetic field configurations~\cite{Christie_2019}. Under those (uncertain) assumptions, the disk winds dominate over the $Y_e \lesssim 0.1$ dynamical ejecta for all of the configurations studied here, though not necessarily by much.

For simulations that yield very low unbound masses (\(\lesssim 0.001\,M_\odot\)), we report these values as nonzero but note that they approach the limits of what can be reliably resolved at our current numerical resolution. 
In the SFHo simulation performed at three resolutions, we find $\sim 25\%$ variations in the unbound mass between different resolutions. This is comparable to the total amount of mass ejected in the SFHo simulation with the lowest black hole spin. In that simulation, the uncertainty on the amount of ejected mass could be $O(100\%)$
While we include these ejecta masses in our analysis, we interpret the smallest values as potentially consistent with zero within numerical error and advise caution in overinterpreting their physical significance.

The distinction between ``red'' and ``blue'' ejecta is important for understanding the nucleosynthesis and electromagnetic signatures of BHNS mergers. Ejecta with low electron fractions ($Y_e < 0.1$) consist primarily of neutron-rich material ejected promptly during the merger due to tidal interactions, leading to high-opacity outflows. In contrast, material with higher electron fractions ($Y_e \sim 0.2 - 0.3$) originates from later-time disk winds, which contain fewer lanthanides and produce lower-opacity outflows \cite{metzger_kilonovae_2019}.

For BHNS systems near the tidal disruption threshold, the resulting kilonovae are expected to be relatively dim compared to events like GW170817. The emission would consist of a weak, infrared transient from the neutron-rich tidal ejecta and a dominant, faster-evolving optical counterpart driven by disk winds~\cite{Barnes:2013wka}. Across all simulations presented here, the total ejecta mass available for $r$-process nucleosynthesis is $\lesssim 0.02M_\odot$. While the neutron-rich dynamical ejecta contribute to $r$-process nucleosynthesis, their low mass suggests a limited role in the overall production of heavy elements \cite{Holmbeck_2023}. 

Generally, the discovery of EM emission from BHNS merger events with a disrupting neutron star has the potential to improve source localization and redshift estimates, and provide more information about the merger process, the origin of heavy elements produced through $r$-process nucleosynthesis \citep{Freiburghaus_1999}, the equation of state of dense matter, as well as the properties of the compact binary itself. As such, it is of utmost importance to accurately predict which BHNS binaries have the potential for neutron star disruption; this will maximize the science returns when modeling these types of binaries.

Building upon the insights gained from this study, future research endeavors could explore additional parameters influencing the behavior of dynamical ejecta in BHNS binaries, such as precessing spins. Furthermore, observational campaigns aimed at detecting and characterizing cbGRBs associated with very low mass disk remnants could provide valuable constraints on theoretical models and computational simulations. Additionally, efforts to refine existing mass-remnant models and to incorporate additional astrophysical and thermodynamic complexities will be crucial for advancing our understanding of compact binary mergers across the electromagnetic spectrum. In that respect, the recent publication of simulations in the $q=3$ regime by multiple research groups (\cite{Izquierdo:2024rbb,Chen:2024ogz}, as well as the results presented here) provides an opportunity to review the assumptions about tidal disruption and ejected matter used in the analysis of GW230529 and in the determination of the probability that BHNS mergers are EM bright~\cite{Abac_2024}. Incorporating our results in that analysis should lead to predicting the production of higher disk masses close to the disruption limit, though the relatively small differences observed here between model and simulations are unlikely to qualitatively change the results of~\cite{Abac_2024} for the likelihood that GW230529 was a disrupting BHNS merger.

These findings also highlight the importance of continued numerical relativity investigations in the parameter space between plunging and disrupting BHNS systems. While the disruption threshold predicted by FNH18 remains broadly reliable, the systematic deviations we observe in disk mass near that limit indicate that a modest recalibration of the fitting formula may be warranted as more simulations at low mass ratio and near-critical spin become available. Expanding such datasets, including cases with precession, mild eccentricity, and more detailed microphysics, will be essential to refine empirical models used by the LVK and to improve predictions for the electromagnetic counterparts of future mixed-binary detections.

\section{Acknowledgements}
T.M., F.F., and M.D. acknowledge the generous financial support from various organizations. T.M. and F.F. acknowledge support from NASA grant 80NSSC18K0565. F.F. acknowledges support from the Department of Energy grants DE-SC0025023 and DEAC02-05CH11231, as well as NSF grant AST2107932. M.D. acknowledges support from NSF grant PHY-2110287. L.K. acknowledges support from NSF grant PHY-2207342 and OAC-2209655. M.S. acknowledges support from NSF grants PHY-2309211, PHY-2309231, and OAC-2209656. L.K. and M.S. also thank the Sherman Fairchild Foundation for their support. 

\bibliography{bibliography}
\bibliographystyle{ieeetr}

\end{document}